\documentclass{ws-procs975x65}

\newcommand{\lsim}{\lower.5ex\hbox{$\; \buildrel < \over\sim \;$}}
\newcommand{\gsim}{\lower.5ex\hbox{$\; \buildrel > \over\sim \;$}}
\newcommand{\lesssim}{\lower.5ex\hbox{$\; \buildrel < \over\sim \;$}}
\newcommand{\gtrsim}{\lower.5ex\hbox{$\; \buildrel > \over\sim \;$}}

\newcommand{\e}{\epsilon}

\newcommand{\ebw}{e_{Bw}}
\newcommand{\nat}{Nature}
\newcommand{\mnras}{MNRAS}
\newcommand{\araa}{ARAA}
\newcommand{\prl}{PRL}
\newcommand{\apj}{ApJ}
\newcommand{\apjl}{ApJ}
\newcommand{\apjs}{ApJS}

\begin{document}

\title{The Collapsar and Supranova Models }

\author{Charles D.\ Dermer}

\address{Code 7653, Naval Research Laboratory\\
4555 Overlook Ave.\ SW,  
Washington, DC 20375-5352 USA\\ 
E-mail: dermer@gamma.nrl.navy.mil}

\maketitle

\abstracts{This rapporteur review summarizes results presented in 
Parallel Session GBT2 (Gamma Ray Burst Theory 2) on the Collapsar and
Supranova Models held 25 July 2003 at the 10th Marcel Grossmann
Meeting on General Relativity, Rio de Janeiro, Brazil.  A central
issue in GRB studies is the process whereby energy is released from
the GRB engine. One scenario is the collapsar model, where the evolved
stellar core promptly collapses to a black hole surrounded by a
massive, intermittently-accreting torus of nuclear density material.
A second scenario is the supranova model, where the first step of a
two-step collapse process leaves behind a rapidly rotating neutron
star stabilized by rotation, which later collapses to a black hole
while making the GRB. In the supranova model, a powerful pulsar wind
lasting days to weeks after the supernova makes distinctive signatures
from the heating of the supernova remnant (SNR) shell, which should be
discovered with Swift. This model also predicts nonstandard reddened
excesses from the cooling SNR due to the range of delays between the
two collapse events, contrary to observations of SN emissions from low
redshift GRBs which favor the collapsar model. The observational basis
for both models is critically reviewed. Problems of the
collapsar/internal shock scenario powered by the Blandford-Znajek
process, and a two-collapse supranova scenario powered by the B-Z
process or the Penrose mechanism, or even Ruffini's pair
electromagnetic pulse, are briefly discussed. In view of the standard
$\gamma$-ray energy upper limit, an external shock model involving
jetted relativistic ejecta powered by an explosive event provides the
most consistent explanation for GRB observations.  An outline of a GRB
model is proposed where the second collapse takes place within minutes
to hours after the primary SN event, during which time loss of
centrifugal support along the rotation axis of the neutron star
provides a relatively baryon-clean polar environment along which a
newly formed, rapidly spinning black hole drives collimated
baryon-dilute outflows.}

\section{Introduction}

The contributions\cite{bot04,laz04} entitled ``The Diagnostic Power of
X-ray Emission Lines in GRBs," by Dr.\ Markus B\"ottcher (Ohio
University, USA), and ``Gamma-Ray Burst Progenitors confront
Observations," by Dr.\ Davide Lazzati (IoA, Cambridge, UK), frame the
current debate. B\"ottcher's article\cite{bot04} summarizes X-ray
observations, and the case is made for a supernova (SN) taking place
weeks to months before the GRB, so that the circumburst medium (CBM)
contains dense ($\gtrsim 10^{10}$ cm$^{-3}$), distant ($\approx
10^{15}$ -- $10^{17}$ cm) SNR shell material whose illumination could
account for the appearance of X-ray features in GRB prompt and
afterglow radiation. These features give important clues about the GRB
environment, and would be compelling if their significance were only
greater. The marginal (3 -- 5 $\sigma$) detections near threshold make
for concern about the reality of the features.  Swift should clear
this issue up, with its $24^\prime\times 24^\prime$ FoV X-ray
telescope reaching $2\times 10^{-14}$ ergs cm$^{-2}$ s$^{-1}$ in a
$10^4$ second observation, with autonomous slewing capability to get
on target within one minute(!).

At the time of this writing (April, 2004), it seems probable that all
GRBs have associated SNe as identified by optically thick photospheric
SN emissions appearing as reddened optical excesses in GRB afterglow
light curves some 10 -- 30 days after the GRB event. In low redshift
GRBs, and most clearly in the case of GRB 030329\cite{hjo03}, the SN
emissions are well fit by SN 1998bw light curves K-corrected and
stretched to match the redshift of the GRB source. These emissions are
powered by trapped shock energy and radioisotope heating (principally
due to the $^{56}$Ni$\rightarrow^{56}$Co$\rightarrow ^{56}$Fe decay
chain formed by explosive nucleosynthesis or accretion disk
winds\cite{psm04}).  The optical follow-on observations indicate that
any delays between collapses to a neutron star and black hole (or
strange star) last for less than a few days. This places constraints
on the angular momentum and mass at formation, and the equations of
state of GRB progenitors\cite{sha04,wmd04}.

Lazzati\cite{laz04} argues that the issue of SN excesses in late (10
-- 30 days) time optical afterglow light curves is settled, and he
examines the low redshift $(z =0.17)$ GRB 030329 in detail. As yet, SN
excesses have not been observed from a GRB which also displays X-ray
features. One possibility is that X-ray features will be detected at
high significance by Swift from a GRB with a nearly coincident (within
$\lesssim$ a few days) SN, in which case the original supranova
model\cite{vs98}, with delays between the two collapse events ranging
from weeks to months, is ruled out. If Fe features are firmly
detected, then this might imply ejection of material from an Fe core
(which then does not explain what powers the SN Ic light curve).
Another possibility is that the GRBs displaying Fe features form a
disjoint set from those which display SN emissions. In this case, two
separate population tracks for GRBs which produce SN emissions and
those which produce X-ray line features from the illumination of a
distant reprocessor are implied\cite{wmd04}. It is also possible that
the X-ray features are artifacts of analysis.

As is apparent, the observational situation, particularly with regard
to the X-ray emission lines and absorption features and edges in GRB
spectra, is murky. Then what about the theoretical situation?

The relativistic blast-wave external-shock model with both forward and
reverse shock emissions is generally accepted as the basis to interpret
afterglow observations, and even to apply during the prompt phase in
those GRBs which display enhanced optical reverse shock radiation.  At
least, this model provides a serviceable tool to interpret radiation
fluxes and to derive physical and environmental parameters ($\epsilon_e$,
$\epsilon_B$, $n_{CBM}$, and to distinguish uniform vs.\ windfed CBM).
The radiation processes that make $\gamma$-rays during the prompt
$(\approx t_{90}$) phase is an open question, with external and
internal shock models each proposed. An external shock model, powered
by a single explosive event, is favored on the basis of the data, as
argued below.

In this article, supranova-model predictions of signatures of SNR
shells heated by plerionic emissions emitted by a newly born, rapidly
rotating and highly magnetized neutron star are presented. The heating of
the SNR, and the interaction of the GRB emissions with this shell of
material imply many predictions for the supranova model. We critique
the collapsar and supranova models and argue that they are both
inconsistent with the data (even leaving aside the X-ray
observations). The observations do not (yet) show the predicted
plerionic signatures of a supranova model, nor do they imply standard
reddened emissions at late times if a fraction of GRBs have delays of
weeks to months between the two collapse events (that fraction, and
the validity of the original supranova model, continues to decline in
significance as more observations of reddened excesses
accumulate). The collapsar model cannot explain the prompt GRB
emissions, durations, efficiencies, nor the standard maximum-energy
reservoir result. It predicts a wind-type CBM, which is only seen on
rare occasion\cite{pri02}. The baryon-dilute jet must be able to
penetrate the massive stellar envelope without substantial mixing and
reach $\Gamma$ factors $\gg 100$.

Some new ideas are needed. One possibility is that energy is
dissipated by general relativistic effects within the ergosphere of a
charged, rotating black hole as it is formed, rather than extracted
when a magnetized disk is accreted onto a rotating black hole, as in
the collapsar model\cite{mwh01}. Numerical simulations of core
collapse of massive stars show that material can be evacuated along
the rotation axis of the stellar core, leaving a relatively
baryon-clean environment\cite{bur04}. Within a few minutes to 1 -- 2
days of the SN, a second collapse would drive a jet in the poleward
direction of the newly formed, rapidly spinning black hole to make the
GRB.  This model would make jet formation more feasible than in a
collapsar scenario where nearly baryon-pure ejecta, whether in the
form of a pair fireball or a Poynting-flux dominated flow, has to
``drill a channel" through a massive stellar envelope and avoid the
processes that are so effective at mixing $^{56}$Co in SN 1987A,
namely strong Rayleigh-Taylor mixing instabilities along the jet head
and Kelvin-Helmholtz instabilities along the jet spine and in the
Rayleigh-Taylor fingers.

Such a scenario does not depart from the well-established connection
of long-duration GRBs with high-mass stars born in star-forming
galaxies that end their lives as Type Ic SNe. Indeed, rapidly spinning
Wolf-Rayet progenitors stars are the most likely GRB progenitors, as
in the collapsar\cite{woo93} and some versions (though not
all\cite{vs99}) of the supranova models. The main difference is in
the nature of the collapse process and the origin and meaning of the
prompt GRB emissions, with important implications for the physics of
stellar evolution and black-hole formation. Whether the GRB is formed
by an engine or explosion is important for understanding the physics
of black hole formation. Here we argue for an explosive event (cf.\
Ref.[\cite{kul04}]).

Such a model brings into focus the ``standard-energy reservoir
result''\cite{fra01} or, more accurately, the ``standard-energy upper
limit''\cite{bfk03}, and avoids many unpleasant issues that are often
avoided in discussions of the collapsar scenario, such as low
efficiency and absence of curvature signatures that are expected from
colliding shells in the internal/external scenario, and the frequent
presence of a rather uniform CBM as inferred from afterglow model fits
to the external shock emissions\cite{pk01}.  I also comment on current
models for energy extraction from a rotating black hole, including the
Blandford-Znajek effect and the Penrose mechanism.

Both the collapsar and supranova models have difficulties to explain
the GRB phenomenon. We\cite{dm99,dm03} have argued that the pulses in
a GRB light curve are flashes made when a relativistic blast wave
intercepts density inhomogeneities in the CBM.  Delayed low-level
radiations from fallback onto the newly formed black hole are possible
(and this could help explain X-ray features should they turn out to be
real). The GRB light curves are compatible with an explosion or a
magnetar discharge\cite{uso92}. The relativistic ejecta shell can
support a reverse shock structure to explain prompt optical and early
radio afterglows.


Reviews of GRB physics and observations that focus on
observations\cite{pkw00} and theory\cite{zm04}, both from the
perspective of the external shock model\cite{der02} and the
internal/external scenario\cite{piran99,mes02}, can be consulted for
further background.

\section{The GRB/Supernova Connection}

The discovery that the GRBs which are observed today took place at
cosmological distances led to the development of the relativistic
fireball/blast-wave model\cite{zm04} that generalizes the theory of
SNe to cosmic explosions with relativistic ejections. The coasting
Lorentz factors $\Gamma$ of GRB outflows reach values of $10^2$ --
$10^3$, as quantified by $\gamma\gamma$ attenuation
calculations\cite{ls01,der04,rmz04}. Associated with this discovery
are other important observational results\cite{pkw00} that tie the
long-duration ($t_{90}\gtrsim 1$ s) GRBs to high mass stars and thence
to SNe or some subset of SNe:

\begin{itemize}
\item GRB host galaxies have blue colors, consistent with
galaxy types that are undergoing active star formation\cite{djo01,blo02}.
\item GRB counterparts are found within the optical radii and central
regions of the host galaxies, consistent with locations within
star-forming spiral arms\cite{blo99}.
\item In some cases (e.g., GRB 011121), the surrounding medium is better
fit with a wind-fed than uniform CBM\cite{pri02};
\item Lack of optical counterparts in
some GRBs may be due to high column densities and 
extreme reddening from large quantities of gas and
dust in the host galaxy\cite{gw01}, as found in molecular cloud
complexes where massive stars are born. 
\item In other cases, however, the
{\it HETE-2} mission has shown that the ``dark optical GRBs'' are due
to the large range of temporal indices of GRB optical light curves, which
are sometimes very steep and rapidly fading\cite{lam04}.
\item GRBs are associated with nearly contemporaneous Type Ic SNe
in low-redshift bursts where detection of the SN emissions are
possible\ (see Section 4);
\item GRB outflows are in the form
of highly collimated jets. The evidence for beaming is inferred from
achromatic beaming breaks in optical afterglow light curves that occur
when the Doppler cone of the decelerating relativistic outflow is
about the same size as the opening-angle of the jet\cite{sta99}.
\end{itemize}

From the beaming breaks, Frail et al.\cite{fra01} inferred that the
mean solid angle subtended by {\it observed} GRB jets is about 1/500th
of the full sky, so that the mean beaming factor $f_b = 0.2$\%. As a
consequence, there are many hundreds of GRBs events that take place
for every one that is detected. The absolute total energies in
$\gamma$-rays from most GRBs cluster within a factor of $\approx 2$ of
$\sim 1.3\times 10^{51}$ ergs\cite{bfk03}, typical of the kinetic
energy in SN ejecta.  By performing the statistics of GRB sources,
one\cite{der02a} finds that the rate of GRBs in the Milky Way could
reach $\approx 10$\% of the rate of Type Ib/c SNe.

The value of $f_b$ is currently a subject of debate and is in any case
not precisely defined for structured jets. Larger values of $f_b
\approx 0.01$ -- 0.1 are sometimes claimed, which makes the individual
GRB event rarer and more energetic, even if the local time- and
space-averaged power into a star-forming galaxy remains about the same
based on the hard X-ray/$\gamma$-ray output, $\approx 10^{44}$ ergs
Mpc$^{-3}$ yr$^{-1}$. Thus, not all Type Ic SNe harbor a GRB, and the
fraction of Type Ic SNe that host GRB explosions may be as small as
1\%. Nearby GRBs are likely to radiate less energy in $\gamma$-rays
relative to ejecta kinetic energy compared to more distant
GRBs\cite{kul04}. Rapidly fading GRBs also tend to be underluminous in
$\gamma$-ray energy\cite{bfk03}.

These lines of evidence suggest that GRBs are produced by collimated
relativistic plasma ejected during an explosion or an active engine
phase of a subset of Type Ic SNe formed from the collapse of the cores
of high-mass stars.

\section{X-Ray Features and the Supranova Model}

This section critiques M.\ B\"ottcher's contribution\cite{bot04}.  See
his article for more details, including additional references to the
original literature.

X-ray features have now been detected in at least 8 GRBs. Table 1 of
Ref.\ [\cite{bot04}] summarizes these observations, but the most
important results are:

\begin{enumerate}

\item low significance X-ray Fe K feature observed from 
GRB 970508\cite{pir99} in a secondary outburst with Beppo-SAX,
 consistent with an $^{56}$Fe Ly$\alpha$ emission line blue-shifted by
 0.1c from the redshift $(z = 0.835)$ of the GRB source;

\item  low significance X-ray Fe K feature
 observed in GRB 970828\cite{yos01} $(z = 0.958)$ with ASCA, which is
only consistent with the redshift of the GRB host galaxy if the
feature is identified with an Fe K recombination edge in a
nonequilibrium situation;

\item  X-ray emission features observed in the afterglow spectra
of GRB 991216 $(z = 1.00$) with Chandra\cite{pir00}, identified with
H-like Fe and a recombination edge of Fe, along with weaker evidence
for a recombination edge and emission line of H-like S, which
originate in an outflowing moving at $\sim 0.1c$ with respect to the
source redshift;

\item variable and declining Fe absorption lasting for the 
first $\approx 14$ s of the $\gamma$-ray luminous phase of
GRB 990705\cite{ama00} at $z = 0.8435$;

\item  an X-ray feature observed from GRB 000214\cite{ant00} with 
Beppo-SAX, which had no optical transient counterpart or redshift
 measurement, but if identified with Fe K$\alpha$, corresponds to $z =
 0.47$;

\item   multiple high-ionization emission features of S XVI , Si XIV, 
Ar XVIII and Ca XX (but no Fe K$\alpha$ emission line or edge)
observed from GRB 011211\cite{ree02} $(z = 2.14)$ with XMM-Newton (whose
reality has been questioned);

\item  low significance high-ionization Si XIV and S XVI 
blueshifted by $0.12 c$ with respect to the host galaxy redshift (z =
1.254) observed from GRB 020813\cite{but03} with XMM-Newton from a GRB
triggered by HETE-II; and

\item low significance high-ionization S and Si (no Ni or Fe) 
emission features observed with XMM-Newton from GRB 030227 triggered
by the INTEGRAL Burst Alert System\cite{wat03}. The emission lines
appeared in one segment of the observation, and the X-ray spectrum was
lineless in an earlier segment. The redshift was not measured
directly, but equal to $z = 1.32$ and 1.34 for Si XIV and S XVI,
respectively, which when transformed from the emission to the observer
frame implies $z\approx 1.6$.

\end{enumerate}

 These results, taken individually and as a group, indicate that GRBs
take place in environments highly enriched in metals, such as active
star-forming regions.  A highly metal-enriched and dense shell of
material surrounds the sources of GRBs, and this shell is sometimes
moving at speeds of $\approx 0.1c$ with respect to the source. A major
concern is that although the significance is rather marginal, the
amount of line energy is uncomfortably large and approximately
constant, ranging from about $6\times 10^{48}$ ergs per line in GRB
030227 to $\approx 2\times 10^{49}$ ergs for GRB 970508. The line
fluorescence is typically detected over observing time scales of
$\approx 10^4$ s. The line energies in the sample of GRBs have tended
to drop by a factor $\approx 4$ in the later bursts, but this trend
may be misleading.  Another oddity is that although $^{56}$Fe
K$\alpha$ emission was reportedly observed in the first four GRBs,
only emission lines associated with Si, S, Mg, and Ca were seen in
the later GRBs when XMM-Newton was used, which may
however reflect the different capabilities of the respective X-ray
telescopes.
 
Taking the observations at face value leads to some immediate
requirements for models.

If the total $^{56}$Fe K$\alpha$ line energy is $10^{49}E_{49}$ ergs,
then the number of Solar masses in $^{56}$Fe if each ion is
individually illuminated once is
\begin{equation}
{M(^{56}Fe)\over M_\odot}= {10^{49}E_{49}{\rm ~ergs}\over 5F_5
(1~M_\odot/56 m_p)\cdot 13.6(26^2) Z_{26}^2{\rm~ eV}}
\cong {6E_{49}\over F_5 Z_{26}^2} \;,
\label{56Fe}
\end{equation}
where $F = 5 F_5$ is the fluorescence yield.  That's a lot of Iron,
which if made through stellar explosive nucleosynthesis, must already
have decayed ($^{56}$Ni decays through electron
capture\cite{mcg02,mw02}, so the decay timescale could even be longer
than $\approx 100$ days).

Type II SNe are thought to release perhaps $\approx 0.07 M_\odot$ of
$^{56}$Ni, and Type Ia as much as a Solar mass. Even if Type Ic SNe
release $\approx 0.5 M_\odot$ of $^{56}$Ni, multiple ionization episodes
from a dense medium are required to account for the extraordinary
energy and equivalent widths of the emission lines and absorption
features.

Another constraint\cite{vie99,laz01} applies to the size scale of the
emitting region, which is assumed to be moving at most at mildly
relativistic speeds. From the duration $t_{line}$ of the line
observations, given in units of $10^4t_4$ s, then
\begin{equation}
(1-\cos\theta_{shell})R \lesssim {c t_{line}\over 1+z} 
\cong {3\times 10^{14}t_4\over 1+z} {\rm cm}\;,
\label{R}
\end{equation}
where $\theta_{shell}$ is the angle between the fluorescing or
absorbing shell medium and the observer.  If the target material is
moving in the transverse (equatorial) direction at the speed of
$0.1v_{-1}$c, a {\it torus} geometry is natural. This implies that the
maximum duration between a first and second collapse event is $
\lesssim v_{-1}^{-1}$ day, assuming that the ionizing radiations
travel near light speed. Vietri et al.\cite{vie99} and
B\"ottcher\cite{bot00} and Fryer\cite{bf01} consider a model where an
expanding shell materially impacts and shocks a torus of
material. This geometry permits only short time delays between the
shell ejection and the GRB and X-ray line production. The target shell
material could be from an intense progenitor wind formed during the
common envelope phase of a compact object and helium star.

A {\it distant reprocessor} could be $\approx 10^{15}$ -- $10^{17}$ cm
away and contain clumps of highly enriched material with densities
$\gtrsim 10^{10}$ cm$^{-3}$. Variable X-ray absorption is essentially
undetectable from a GRB surrounded by a uniform CBM, even if highly
enriched\cite{bot99}. Extremely clumped ejecta\cite{bfd02} or
blueshifted resonance scattering in a high-velocity
outflow\cite{laz01} can account for the variable absorption in GRB
990705.  In this geometry, a shell of material is illuminated by the
GRB emissions and could originate from a SN taking place months
earlier.  The target material would have locations and size scales
corresponding to requirements of the external shock model for the
prompt phase\cite{dm99}. There are ranges of ionization parameters
that are compatible with the observations for a geometry-dominated
model, though not an engine-dominated (i.e., collapsar)
model\cite{lrr02}.

Considerable heating accompanies multiple recombinations and
reionizations, thereby reducing the recombination rate coefficient, so
this model is criticized to make such a large line energy. Pair
production could enhance the line production efficiency. Other
possibilities includes a funnel geometry\cite{rm00} with a long-lived
photoionizing source, which has not been detected\cite{tgl04}, and a
pair screen\cite{kn02} to provide a local flux of UV/X radiation. The
question of whether photoionization models or thermal models (or both)
apply remains an open question.

Taken together, the earlier ejection of a stellar envelope would be
advantageous to explain the X-ray features, as in the supranova
model\cite{vs98}. This reasoning, given also the results in Chapter 2,
leads to the conclusion that GRBs are a subset of SNe powered by a
two-step collapse to a black hole of a massive ($\gsim $ several
M$_\odot$) core stripped of its H and He envelope. The collapse event
drives a subrelativistic outflow to power the SN, and a relativistic
outflow to power the GRB.  The GRB outflow is likely to be highly
collimated. The excitation of torus material or the illumination of
shell material in the polar direction could produce the line
signatures. Possible stellar progenitor candidates are Wolf-Rayet
stars\cite{woo93}, or merger events involving He
cores\cite{fwh99,bfd02}.

If the X-ray identifications of Fe features are real, then GRB SNe must
yield $\sim 1 M_\odot$ of $^{56}$Fe, which is a challenge for any
model.

\section{Delayed Reddened Excesses and SN Light Curves}

This section critiques and updates D.\ Lazzati's
contribution\cite{laz04}, and summarizes evidence for SN emissions in
the light curves of GRBs and implications for GRB source models based
on the examples of GRBs 980425 and 030329.

GRB 980425, discovered with Beppo-SAX, was a smooth-profiled GRB with
mean duration $\sim 20$ s, and is likely associated spatially and
temporally with the Type Ic SN 1998bw in a spiral host galaxy at a
distance of $\approx 40$ Mpc.  Its apparent isotropic $\gamma$-ray
energy was only $\approx 10^{48}$ ergs.  It displayed very intense
radio emissions, indicative of a related mildly relativistic
outflow. GRB 980425 was probably an off-axis event with a much lower
$\gamma$-ray to kinetic luminosity than in other GRBs. The kinetic
energy is estimated from the radio luminosity using minimum energy
arguments, or by multiplying the X-ray luminosity with time at a fixed
observing time (e.g., 8 hours after the GRB) when the electrons are in
the fast cooling regime and making a correction factor for the number
of ions per electron\cite{fw01}. The kinetic energy inferred from the radio
luminosity of SN 1998bw was $\gtrsim 10^{49}$ ergs\cite{kul98}, and
exceeds $10^{52}$ ergs when modeling the SN light
curve\cite{iwa98}. These energies are similar to the kinetic energy of
the Type Ic SN 2003L, but orders of magnitude greater than the radio
luminosity in the Type Ic SN 2002ap.

Comparison of GRB 980425 with the low redshift ($z = 0.17$) GRB 030329
associated with SN 2003dh is instructive. Based on the jet-break
times, typical beaming-corrected GRB energies have a ``standard"
$\gamma$-ray energy reservoir of $\sim 10^{51}$ ergs\cite{fra01,bfk03}
(see Section 5).  GRB 030329 displayed an early optical jet break at
$\approx 0.4$ d, and a much later 15 GHz jet break at $\approx 15$
d. In contrast, GRB 030329 has a $\gamma$-ray energy $E_\gamma\approx
2\times 10^{50}$ ergs\cite{wil04}, and is therefore underenergetic
compared to a ``typical'' GRB (as also are XRF 020903 and GRB
980425). But in each of these cases, the kinetic energies
$E_{ke}\gtrsim 10^{51}$ ergs\cite{sod04}.  (By contrast, GRB 031203,
at $z\cong 0.1$, which is very similar to GRB 980425, has $E_\gamma
\approx 10^{50}$ ergs and $E_{ke}\approx 2\times 10^{49}$ ergs.)

All this evidence points to a highly structured jet, or a ``spine in
sheath" geometry (as in the case of 3C 273\cite{bah95}) to account for
GRB detectability and phenomenology. A dirty/clean standard top-hat
jet fireball model is too simplified.  Even if one wished to maintain
a unifying approach to GRB observations, then angle-dependent effects
from a structured jet must be considered. In any case, the number of
GRBs hosted by Type Ib/c SNe should be $\lesssim 10$\%.

The discovery that a SN emerges in the afterglow of GRB 030329 tells
us that at least a fraction of SNe host GRBs, and that SN and GRB
explosions are most likely simultaneous\cite{laz04} or, at most,
separated by a few days.  Here summarizes the evidence for delayed
reddened excesses from SN emissions in GRB optical afterglows:

\begin{enumerate}

\item GRB 970228 at $z = 0.695$, which reached a flux at 
$\nu = 3.8\times 10^{14}$ Hz at $\approx 38$ d after the
GRB\cite{rei99,gal00} of $F_\nu \cong 0.87\mu$Jy, corresponding to a
stationary-frame luminosity of $4\pi d_L^2 \nu F_\nu$(t = 25 d) $\cong
10^{43}$ ergs s$^{-1}$ ($d_L = 1.4\times 10^{28}$ cm), about the same
as the SN 1998bw template.  Reichart\cite{rei99} argues that this
emission signature cannot be explained with a dust echo.

\item GRB 980326 with unknown redshift\cite{blo99a}. 
The peak of the light curve at 6558 \AA (R-band is 7000 \AA, 1.8 eV,
$\nu = 4.3\times 10^{14}$ Hz) appears at $\sim 20$ days and has a peak
flux density $\approx 0.4\mu$Jy, or a $\nu L_\nu$ power of $\approx
10^{43}$ ergs s$^{-1}$ if $z \sim 1$.

\item GRB 980425 $(z = 0.0085$, or $d \approx 40$ Mpc) 
associated with the Type Ic SN 1998bw\cite{gal98,pia00}.  With a peak
absolute luminosity $\approx 1.6\times 10^{43}$ ergs s$^{-1}$, it
requires $\sim 0.7 M_\odot$ of $^{56}$Ni to power the light
curve\cite{iwa98}.

\item GRB 980703 at $z = 0.9661$,  consistent with a SN 1998bw-type
excess at $\approx 10$ -- 20 days after the GRB, or no 
SN\cite{hol01}.

\item GRB 000418 at $ z = 1.119$, which had R-band emission exceeding
$\approx 1~\mu$Jy 40 days after the GRB. If due to jetted afterglow
emission, this flux would correspond to a $\nu L_\nu$ luminosity
$\approx 3\times 10^{43}$ ergs s$^{-1}$ that could conceal a SN
excess\cite{ber01}.

\item GRB 000911 with a host galaxy redshift of $z = 1.06$ and 
optical emission with luminosity $L \sim 1.5\times 10^{43}$ ergs
s$^{-1}$ after $\approx 1$ week.  The addition of a SN 1998bw-type
excess improves the fit, and an alternative model of a shock
interacting with WR progenitors involves several free
parameters\cite{laz01a}.

\item GRB 011121 at $z = 0.362$ with a distinct reddened excess
appearing at $\approx 20$ days after the GRB and reaching a $\nu L_\nu$
power of $\approx 2\times 10^{42}$ ergs s$^{-1}$ at 7020\AA. This
delayed emission is well fit by a SN1998bw template emission, and the
agreement improves if GRB 011121 exploded $\sim 3$ -- 5 days after the
SN\cite{blo02a,pri02}. 

\item GRB 010921 at $z = 0.451$, which showed no SN to a limit 
1.33 magnitudes fainter than SN 1998bw at the 99.7\% confidence
limit\cite{pri03}.

\item GRB 020405 at $z = 0.690$, which showed a delayed reddened 
excess consistent with a SN 1998bw emissions dimmed by 0.5
magnitude\cite{pri03a,ber03}. The radio data are not consistent with a
density profile $\propto r^{-2}$.

\item GRB 020410, redshift unknown, that exhibited an optical
rebrightening one week after the GRB consistent with SN emissions
from a nearly coincident SN\cite{lev04,nic04}.

\item GRB 030329, associated with SN 2003dh at $z = 0.1685$, with 
a highly variable optical component\cite{sta03,lip04,hjo03}.

\item The X-ray flash XRF 031203 at $z = 0.1055$, associated with SN 2003lw,
which can be represented by SN1998bw brightened by 0.55
mag\cite{tho04}.

\end{enumerate}

These observations show that in many cases, GRBs and XRFs are
accompanied by a nearly contemporaneous SN which is spectrally much
like the Type Ib/c SN 1998bw that is temporally and spatially
associated with GRB 980425.

A collapsar model GRB must have a progenitor with stellar photospheric
radius $\lesssim 10^{11}$ cm, which limits allowed progenitors to
massive stars that have lost their H and He envelopes, such as
Wolf-Rayet stars or close binary systems\cite{mat03}.  They must keep
the ejecta extremely baryon-free by pushing away $\sim 0.01 M_\odot$
while retaining a baryon purity $\lesssim 10^{-3} M_\odot$ mixed in
with the explosion. A model for jet emission must explain the Amati
relation\cite{ama02}, where the $\nu F_\nu$ peak energy $E_{pk}
\propto E^{1/2}_{iso,\gamma}$, where $ E_{iso,\gamma}$ is the apparent
isotropic energy release.

Although Lazzati had argued that the bumps and wiggles in the light
curve of GRB 021004\cite{laz02} were due to density enhancements in
the CBM, he abandons that model for a model with episodes of delayed
injection to explain the complex optical light curve of GRB
030329\cite{lip04}. He claims that `any event that takes place over
the entire fireball surface' will produce deviations in the afterglow
light curve `on timescales $\delta t\sim T$, where $T$ is the moment
in which the deviation begins.' '' This argument does not apply when
the size scale of the density irregularities is much smaller than
$\Gamma^{-1}$, which is the most interesting regime from the point of
view of an external shock model. Besides the complexity expected in
the CBM\cite{wl00}, special relativistic effects\cite{dm99,dm03}
enhance emissions from shell-cloud interactions for clouds with
$\theta \ll 1/\Gamma$.

The problem of the interpretation of the optical polarization curve of
GRB 030329 may be solved if complex behavior in its optical afterglow
light curve is a result of localized blast-wave/cloud interactions in
an external shock model.

\section{The Standard Energy Reservoir and the Collapsar Model}

Because of the potential importance of this finding, the ``standard
 energy reservoir'' result is summarized before enumerating the
 internal contradictions of the collapsar model.

The standard $\gamma$-ray energy release from a GRB holds at least as
an upper limit within a factor of $\approx 2$, and was originally
quoted as $E_\gamma\approx 3\times 10^{51}/(\eta_{\gamma}/0.2)$
ergs\cite{fra01}, with the efficiency for $\gamma$-ray production
compared to the total energy in the outflow given by
$\eta_{\gamma}\approx 0.2$. The absolute $\gamma$-ray energy from a
GRB is given by $E_\gamma = f_b E_{iso,\gamma}$, where the beaming
factor $f_b = 1-\cos\theta_j$, $ E_{iso,\gamma} = 4\pi d_L^2
F_\gamma/(1+z)$, and the jet-opening half-angle given by\cite{fra01}
\begin{equation}
\theta_j = 0.057\left({t_j\over {\rm 1~ day}}
\right)^{3/8}\left({1+z\over 2}\right)^{-3/8}
\left[{E_{iso,\gamma}\over 10^{53} {\rm~ergs}}\right]^{-1/8} 
\left({\eta_\gamma\over 0.2}\right)^{1/8}
\left({n\over 0.1{\rm~cm}^{-3}}\right)^{1/8}\;.
\label{thetaj}
\end{equation}
Note the weak dependencies on the unknown densities and efficiencies,
which can be further constrained by afterglow modeling and comparison
of the measured $\gamma$-ray energy with the kinetic energy of the
flow.

The standard $\gamma$-ray energy reservoir result is now superseded by
observations\cite{bfk03} of low $E_\gamma$ outliers GRBs 980425,
030329, and 031203 which are underluminous, implying that the
standard $\gamma$-ray energy is an upper limit and that there are
biases favoring the detection of distant GRBs with maximum energy
release.

This result is quite difficult to explain in the context of a
collapsar scenario, where the relativistic jet passes through a
massive dense stellar envelope.  A ``large diversity in any
accompanying SN components of GRBs is expected,'' and only a tiny
fraction of a Solar mass of baryons mixed into the explosion would
make for a considerably different amount of energy in $\gamma$-rays,
so the ``apparent constancy of the gamma-ray energy release is even
more mysterious\cite{blo02a}.''

Some other particulars which an internal shock/collapsar model must answer
to be viable are:
\begin{enumerate}
\item How is it possible for pulses in GRB blast waves to have high efficiency
in an internal shock model and also to have approximately constant
(within a factor of 2 -- 3) values of peak photon energy $E_{pk}$ in
different pulses of the same GRB in spectra that extend to hundreds of
MeV?
\item If internal shells have such widely different efficiencies, then
how is the standard $\gamma$-ray energy reservoir result explained?
\item Why are curvature effects from spherical GRB blast waves, as
found in an internal shock scenario, rarely if ever seen in GRB pulse
profiles\cite{br01,der04a}?
\item How is it possible for a jet to pass through a massive stellar envelope
of several light seconds width and create pulses that are fractions of
a second long?
\item If the collapsar model is a failed SN, that is, there is no associated
SN, then how is the enormous amount of $^{56}$Ni produced to
power the light curve\cite{mwh01}?
\item Why have there been no detailed comparisons 
between temporal and spectral characteristics of GRB pulses, or even a
smooth fast-rise, slow decay light curve, to show whether an internal
shock scenario with high efficiency can explain GRB data?
\end{enumerate}

An external shock model with a single relativistic blast wave can
simply answer these questions:

\begin{enumerate}
\item An external shock model is much more efficient than 
an internal shock model\cite{dm03}, and approaches 100\% efficiency
for $\gamma$-ray production from accelerated electrons when the CBM is
sufficiently dense. The values of $E_{pk}$ are relatively constant
because the relative Lorentz factor in the interactions is about the
same for interactions within the Doppler beaming cone.
\item The most energetic GRBs are those where the 
largest fraction of the available kinetic energy is dissipated in the
form of $\gamma$ rays.
\item Curvature effects are not expected to be seen in an external shock model
where the blast wave intercepts inhomogeneities on angular scales
smaller than $\approx 1/\Gamma$.
\item The jet originates from a bare collapse into a medium that has already
been evacuated of much of the dense material.
\item A two-step collapse process would first form a SN with a large
amount of $^{56}$Ni production, and would later collapse to a black
hole to produce the GRB.
\item Comparisons of pulse profiles, time histories, and phenomenology with
an external shock model are favorable\cite{dbc99,bd00}.
\end{enumerate}

Zhang and M\'esz\'aros\cite{zm04} review the debate.

\section{The Heated SNR Shell in the Supranova Model}

Any viable GRB model must be able to explain the increasing evidence
for supernova-like emissions in the optical afterglows within the
context of a model that explains the highly variable pulses in the
prompt phase of a GRB.  The evidence for X-ray features in prompt and
afterglow GRB spectra indicates that high density, high metallicity material
near GRB sources.

Is it possible to construct a model to explain at the same time the
X-ray features and the delayed reddened excess emissions?  Could a SNR
shell that is illuminated and heated by a pulsar wind (PW) and cools
following the GRB event make emissions that explains the delayed
reddened excesses seen from GRBs?

We\cite{der02b} have tried to construct a model within the context of
the two-step collapse process in the supranova scenario.  This model
has a number of attractive features: the period of activity of a
highly magnetized neutron star preceding its collapse to a black hole
can produce a pulsar wind bubble consisting of low density, highly
magnetized pairs\cite{kg02,igp03}, in accord with afterglow model
fits\cite{pk01}.  The heating of the SNR shell by the X-ray
synchrotron radiation from the pulsar wind electrons could provide an
important source of ionizing radiation. A cooling SNR shell could
produce the excess reddened emissions.

We have found, however, that such a solution for the reddened excesses
requires fine-tuning and, if not fine-tuned, predicts a wide range of
cooling emissions that have not yet been observed. (It is important,
however, to measure the spectra of SN emissions from nearby GRBs to
determine how similar they are to SN 1998bw.) Thus the original
supranova scenario appears not to be compatible with observations of
SN emissions if all GRBs, as seems probable, display rather
uniform SN emissions in GRB optical afterglows.

If the delayed reddened excesses in GRBs cannot be explained by the
supranova model, it is ruled out. A simple derivation of the emissions
from a SNR shell heated by a PW is presented in the Appendix.

The results of that analysis are displayed in Fig.\ 1, which shows the
 total $\nu L_\nu$ flux composed of synchrotron, SSC, and thermal
 components.  Standard parameters are used,as defined in the Appendix.
 The heavy solid and dashed curves correspond to standard parameters
 with $t/t_{sd} = 0.1$ and 0.01 as labeled, where $t_{sd}$ is the
 magnetic dipole spin-down timescale, the crossing (or dynamical)
 timescale of the shell $t_{dyn} = 32.4$ and $ 3.2$ days, and $T_{eff}
 = 1700$ and $5420$ K, respectively ($t_{sd} = 3240$ d). This system
 is in accord with the standard supranova model and might explain the
 delayed excess emission in GRB 011121, but requires fine-tuning.  The
 set of three dotted curves employ standard parameters, except that
 $B_{12} = 4$, and $t/t_{sd} = 1, 0.1,$ and 0.01 as labeled ($t_{sd} =
 202$ d), with $t_{dyn} = 20.2, 2.0$, and $ 0.2$ days ($= 17$ ks), and
 $T_{eff} = 3760, 11900,$ and $37630$ K, respectively.

The intense PW X-ray synchrotron emission provides a source of
ionizing photons throughout the SNR shell that would decay on the
timescale $t_{dyn}$ (the line flux would decay on $\approx 2 t_{dyn}$
due to emissions from the far side of the shell). The $B_{12} =4,
t/t_{sd} = 0.01$ case might correspond to a GRB 011211-type
system. The PW radiation field provides an additional source of
ionizing photons in the supranova model,
 and would be energetically important for this case if the typical
energies of electrons injected by the pulsar wind is
$\gamma \cong 10^4$. A varying ionization flux impinging on an
expanding SNR shell will produce a characteristic kinematic signature
of shell illumination.

\begin{figure}[t]
\vskip-2.0in
\centerline{\epsfig{file=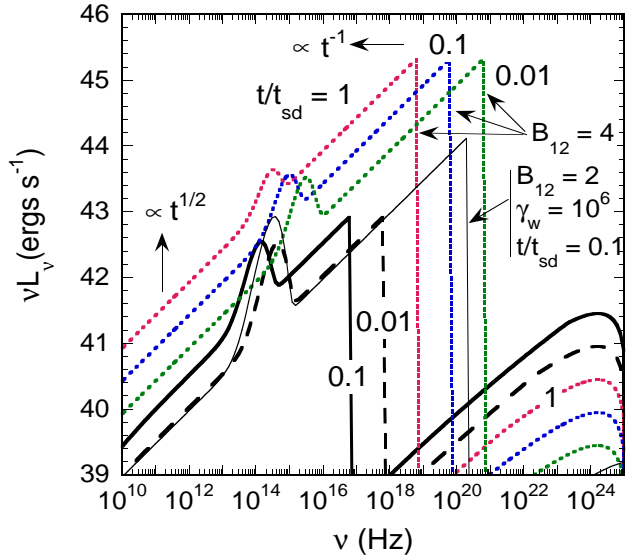,width=13.0cm,height=16.0cm }}
\vspace{-1.7in}
\caption{ Total model spectral energy distribution
composed of synchrotron, synchrotron self-Compton, and shell thermal
emission from a pulsar wind and wind-heated SN shell. Standard parameters,
given in the Appendix, are used, except where noted in the legends. 
\label{fig:KASCADE}}
\end{figure}

The thin solid curve is the result for standard parameters except that
$B_{12} = 2$, $\gamma_w = 10^6$, and $t = 0.1 t_{sd} = 81$ d.  For
this case, $t_{dyn}=8.1$ d and $T_{eff} = 4360$ K. It is apparent that
a range of parameter could make quasi-thermal emission between
$\sim 10^{14}$ -- $10^{15}$ Hz at the level of $\sim 10^{42}$ -- $10^{43}$
ergs s$^{-1}$ some tens of days after the GRB event, but only if 
accompanied by intense heating from the plerionic synchrotron radiation.

Thus, a delayed reddened excess from the cooling shell may be seen,
for plausible parameters, some tens of days after the GRB even if the
GRB takes place months after the SN.  The PW heating is in addition to
any heating due to the Ni$\rightarrow$ Co $\rightarrow$ Fe chain, and
may relax the need to have $\gtrsim 0.5 M_\odot$ of $^{56}Ni$, as
required to explain the unusually bright light curve of SN
1998bw\cite{iwa98}. Whether the heated shell looks like the observations
of reddened emissions in the optical emissions of GRBs will require
better observations of Type Ic SNe and low-redshift GRBs, and better 
studies of this system.

Should this model be correct, then GRBs would exhibit a bright hard
X-ray emission feature corresponding to the plerionic component that
could be detected with INTEGRAL or Swift from nearby $z \lesssim 0.1$
-- 0.3 GRBs when the PW synchrotron flux radiates $\gtrsim 10^{44}$
ergs s$^{-1}$ at X-ray energies. A nonthermal optical emission
signature would accompany the reddened excess, with a relative
intensity that depends sensitively on the energies of the PW electrons
accelerated. A hard X-ray sky survey instrument such as EXIST could
monitor for the PW nonthermal emission preceding aligned and off-axis
GRBs. The X-ray synchrotron radiation provides a source of external
radiation to enhance photomeson production by energetic hadrons over
the standard GRB model\cite{da03}.

The supranova model makes a number of other predictions:
\begin{enumerate}
\item Compton-scattered $\gamma$ radiation from GRB electrons scattering
photons from the intense plerionic radiation field\cite{igp03,gg03};
\item High-energy neutrinos from the interactions of accelerated
protons with the SNR shell, and due to photomeson production
with the external radiation field\cite{da03,gg03a,rmw03}.
\item $\gamma$-ray emissions from the newly-formed, rapidly 
rotating pulsar\cite{wmd04} prior to the second collapse to a black hole.
\end{enumerate}

\section{Energy Release from a Newly Formed, Rapidly Rotating Black Hole}

At the heart of the GRB debate is the process that liberates an
enormous amount of energy over a very short time in the form of a
collimated plasma outflow with very small baryon contamination. Though
some models have considered highly magnetized pulsar
emissions\cite{uso92} or (as was discussed in the Session) collapse
to quark matter\cite{lug04}, it is generally assumed that GRBs are
powered by the birth of a newly formed black hole. Until the question
of the power source of GRBs is solved, problems about the origin of
$^{56}$Ni, which is supposed to power the reddened excesses observed
in Type Ic SNe---whether from explosive nucleosynthesis or disk
winds\cite{psm04}---cannot be solved.

Here we assume that a GRB is powered by the birth of a newly formed,
rapidly spinning black hole. In addition to the enormous gravitational
energy available during the collapse process, a spinning black hole
can be formed with $\gg 10^{52}$ ergs of rotational energy.  Ruffini
and colleagues\cite{brx01} argue that polarization of the vacuum can
produce an e$^+$-e$^-$ pair-electromagnetic pulse of enormous energy
that, given an external shock model, could explain the light curves of
GRBs. The problem here is that the pair pulse is isotropic, contrary
to inferences from observations that GRB ejecta are highly
collimated. Moreover, unless the collapse to a black hole takes place
well after the SN, it would be difficult for an isotropic
explosion to break through the stellar envelope.

Tapping the spin energy of the newly formed black hole seems a better
bet for a GRB model. Advocates of a collapsar model generally invoke the
Blandford-Znajek process where an induced voltage, formed when
the intense accretion-disk magnetic field is shorted out in the
ergosphere of the spinning black hole, accelerates particles. For a
sufficiently massive and highly magnetized accretion disk surrounding
a black hole with spin parameter $a = Jc/GM^2\rightarrow 1$, adequate
power can be injected to explain a GRB\cite{mwh01}. Though this is the
most popular model for GRBs at the present time, quantitative
explanations of temporal and spectral features of GRB-like pulses, the
constant energy-upper limit, $^{56}$Ni production, the escape from the
stellar envelope of a lage $\Gamma$ outflow, the mean duration of GRB
activity, etc.\ (see Section 5), remain unsolved. Another problem is
that simulations of collapse to a black hole do not leave behind a
sufficiently massive torus to power a GRB\cite{sha04,sha00}. Such
simulations do not give much support for a supranova scenario, either,
because differential rotation induces toroidal Alfv\'en waves that can
magnetically brake and destabilize the neutron star, thereby causing
collapse within hundreds of seconds.

Another possibility is that the Penrose process, involving either
Penrose Compton scattering or Penrose pair
production\cite{ps77,lk78,wil95}, extracts energy from the newly
formed, rapidly rotating black hole. Particles on retograde orbits in
the ergosphere can extract rotational energy through decay or
scattering. It is not clear, however, that retrograde scatterers can
be formed in the dynamically forming black-hole, nor is it clear that
the efficiencies of this process are adequate to power a GRB.

Given the uncertainties in our knowledge of the fundamental GRB energy
release process (``a black hole within a black box''), it might be
best to speculate on a viable GRB model from the observations, as
summarized earlier.

\begin{enumerate}

\item The delayed reddened excesses are indeed due to SNe, and the progenitor
star of a GRB collapses to a neutron star to form a Type Ic SNe,
forming $^{56}$Ni through explosive nucleosynthesis.

\item The delays between collapses from a neutron star to a black hole
last from minutes to days rather than from weeks to months. This 
means that X-ray features associated with Fe are mistakenly identified,
though the signatures of S, Si, and Ca could still be real.

\item During the period between the two collapse events, powerful
pulsar winds could evacuate a channel along the rotational axis of the
progenitor star and the pulsar, where the overlying material has the
smallest column density.

\item A single, highly collimated explosive event during the 
collapse of the neutron star to a rapidly rotating black hole
could eject a relativistic outflow to produce the GRB prompt
and afterglow emissions through external shocks with the CBM.

\end{enumerate}

Swift will be crucial to test this and other GRB models by providing
 better measurements of the X-ray features, by discovering many nearby
 GRBs from which delayed reddened emissions can be precisely measured,
 and by searching for signatures of a heated SNR shell.

\section*{Acknowledgments}
The work of CD is supported by the Office of Naval Research and {\it
GLAST} Science Investigation Grant No.\ DPR-S-1563-Y.

\appendix

\section{Pulsar Wind Physics}

A pulsar wind (PW) and pulsar wind bubble (PWB) consisting of a
quasi-uniform, low density, highly magnetized pair-enriched medium
within the SNR shell is formed by a highly magnetized neutron star
during the period of activity preceding its collapse to a black
hole\cite{kg02}.  The interaction of the PW with the shell material
will fragment and accelerate the SNR shell, and the PW emission will
be a source of ambient radiation that can be Comptonized to gamma-ray
energies\cite{igp03,gg03}.

The nonthermal leptonic PW radiation provides a source of hard
ionizing radiation to heat and photoionize material in the SNR shell,
which can be described in terms of a partial covering model as a
result of the strong PW/SNR interaction. This radiation heats the
shell, and the escaping thermalized radiation is seen as the delayed
reddened optical excesses in GRB optical afterglow light curves.  The
source of ionizing nonthermal radiation, which is extinguished when
the GRB occurs, would provide a nonthermal precursor to a GRB, though
at low flux.

A very rapidly rotating, strongly magnetized neutron star is formed in
the first step of the supranova scenario. For a neutron star with
surface polar field of $10^{12}B_{12}$ G, a rotation rate of
$10^4\Omega_4$ rad s$^{-1}$, and a radius of 15$r_{15}$ km, the
radiated power $L_0\cong 7\times 10^{44}\; B_{12}^2 r_{15}^6
\Omega_4^4\; \; {\rm ergs~s}^{-1}$.  The quantity $E_{rot} = {1\over
2} j (GM^2/ c^2)\Omega\;\cong\;2\times 10^{53}\;j_{0.5}M_3^2
\Omega_4\; {\rm ergs}$ $\equiv 10^{53} E_{53}^{rot}$ is the initial
rotational energy of the neutron star, expressed in terms of the
neutron-star mass $M = 3 M_3 M_\odot$ and the dimensionless angular
momentum $j_{0.5} = GM^2/2c$, which remains roughly constant during
collapse\cite{vs98}. The spin-down time $t_{sd} = (E_{rot}/ L_0) \cong
2.8\times 10^8 \; j_{0.5}M_3^2 (B_{12}^2 r_{15}^6 \Omega_4^3)^{-1}$ s.

The torque equation, $\dot \Omega = - K\Omega^n$, where the braking
 index $n = \ddot\Omega\Omega/\dot\Omega^2$, can be solved to give
 $\Omega(t) = \Omega_0(1+t/t_d)^{-1/(n-1)}$, provided that $K$ remains
 constant\cite{ato99}. (For the Crab, $n = 2.5$.) We assume that the
 spin-down power is radiated as a PW in the form of particle and
 Poynting flux, so that the wind luminosity $L_w \cong L_{sd}
 =\;(\xi_e+\xi_p+\xi_B)L_{sd} = L_0/(1+t/t_{sd})^k$, where $k =
 (n+1)/(n-1)$, and we have divided the wind power into leptonic
 (``e"), hadronic (``p"), and electromagnetic (``B") fractions
 $\xi_i,$ $i=e,p,B$.

In the supranova model, rotational stability is lost due to spindown,
 and the neutron star collapses to a black hole to produce a GRB at
 time $t_{coll}$. Here we assume $t_{coll} \lesssim t_{sd}$, so that
 for $k > 1$ the wind energy $E_w(t) = \eta\;(k-1)^{-1}L_0
 t_{sd}\;[1-(1+(t/ t_{sd})^{1-k}]\rightarrow \eta L_0t $ when $t
 \lesssim t_{sd}$, where the parameter $\eta$ roughly accounts for the
 escape of wind energy from the SNR shell as well as losses of wind
 energy due to radiation, and is assumed here to be constant with
 time.  The equation for the SNR shell dynamics in a spherical
 approximation for the SNR shell is $M_{SNR}\ddot R = 4\pi R^2
 p_w(t)$, where $M_{SNR} = m_{SNR}M_\odot$ is the SNR shell mass, the
 wind pressure $p_w(t) \simeq E_w(t)/3V_{pwb}$, the PWB volume
 $V_{pwb} =4\pi R^3(t)/3$, and the PWB radius $R(t) = \int_0^t
 dt^\prime \; v(t^\prime )$. This equation can be solved\cite{gg03} to
 give
\begin{equation}
R(t) = v_0 t (1+ \sqrt{{t\over t_{acc}}})\;,\;{\rm where~}\;\;t_{acc}
\;= {3 M_{SNR}v_0^2\over 4\eta L_0}\;\cong 1.9\times 10^7\; 
{m_{SNR}\beta_{-1}^2\over \eta B_{12}^2 r_{15}^6 \Omega_4^4}\;\;{\rm s}\;
\label{R(T)}
\end{equation}
It follows that 
$t_{acc}/ t_{sd}\;= E_{ke}^{SNR}/ \eta E_{rot}\simeq 0.1 E_{52}^{SNR}/
\eta E_{53}^{rot}$, where ${1\over 2} M_{SNR} v_0^2 \cong 9\times
10^{51} m_{SNR}\beta_{-1}^2$ $ \equiv 10^{52}E_{52}^{SNR}$ ergs is the
SNR shell kinetic energy, and $v_0=0.1\beta_{-1}c$ is the SNR shell
coasting speed.  If $\eta \cong 1$, then $t_{acc} \ll t_{sd}$ for
standard values, and shell acceleration must be considered. A highly
porous shell has $\eta \ll 1$. This will occur for strong clumping of
the ejecta, which takes place on the Rayleigh-Taylor
timescale\cite{igp03} $t_{\rm RT} \simeq \sqrt{x/\ddot R}$, where the
characteristic clumping size scale is $x \equiv f R$. Hence $(t_{RT}/
t_{acc})\;= f^{1/2}\; (t/ t_{acc})^{3/4}$ in the regime $t \lesssim
t_{acc}$. This shows that small-scale ($f \ll R$) clumping will occur
when $f \ll \sqrt {t/t_{acc}}$. Only a detailed hydrodynamic
simulation can characterize the porosity of the shell due to effects
of the pulsar wind, which $\eta$ parameterizes.  Here we assume, in
contrast to the picture of Ref.\ [\cite{kg02}], that a large fraction
of the wind energy escapes the SNR shell and $\eta \sim 0.1$, so that
shell acceleration can be neglected. This effect also causes the shell
to become effectively Thomson thin much earlier than estimated on the
basis of a uniform shell approximation, except for high density clouds
with small covering factor.  A general treatment is needed to treat
shell dynamics, including travel-time delays of the pulsar wind and
deceleration of the SNR shell by the surrounding medium.  When
$t_{coll} \lesssim t_{acc},t_{sd}$, as assumed here, $R \cong v_0 t$.

The strong PW provides a source of nonthermal leptons, hadrons, and
electromagnetic field. Dominant radiation components considered here
are leptonic synchrotron and SSC radiation, and thermal emission from
the inner surface of the SNR shell which is heated by the nonthermal
radiation.

\subsection{Wind Synchrotron Radiation}

The volume-averaged mean magnetic field $B_w$ in the SNR cavity
powered by the PW is obtained by relating the wind magnetic field
energy density $u_B = B_w^2(t)/8\pi = \ebw u_w(t) = 3 \ebw \eta L_0
t/4\pi R^3$ through the magnetic-field parameter $\ebw$, also assumed
to be constant in time. Thus
\begin{equation}
B_w(t) = {{\cal B}\over t}\;,\;{\rm where~}\; {\cal B} = 
\sqrt{6\ebw L_0 \eta\over v_0^3}\cong 4\times 10^8 \sqrt{\ebw \eta}
\;({B_{12} r_{15}^3 \Omega_4^2 \over \beta_{-1}^{3/2}})\;\; {\rm G-s}\;.
\label{Bwt}
\end{equation}
Let $\gamma_w = 10^5 \gamma_5$ represent the typical Lorentz factors
of leptons in the wind zone of the pulsar. This value should remain
roughly constant when $t \lesssim t_{sd}$. A quasi-monoenergetic
proton/ion wind may also be formed, though we only consider leptonic
processes here, and furthermore do not treat nonthermal particle
acceleration at the PW/SNR shell boundary shock.  When synchrotron
losses dominate, the mean Lorentz factor of a distribution of leptons
with random pitch angles evolves in response to a randomly oriented
magnetic field of mean strength $B$ according to the expression $-\dot
\gamma_{syn} = (\sigma_{\rm T} B^2/ 6\pi m_e c)\gamma^2$, giving
\begin{equation}
\gamma_w = [{1\over \gamma} - T_B\;({1\over t_i} - {1\over t})]^{-1}\;\;,
\label{gammaw}
\end{equation}
where $t_i$ is the injection time, $\gamma_i = \gamma_w$ is the
injection Lorentz factor, and
\begin{equation}
T_B \gamma_w = {\sigma_{\rm T}{\cal B}^2\over 6\pi m_e c}\;
\gamma_w\;\simeq\; 2\times 10^{13} \;\ebw\eta\; 
({B_{12}^2 r_{15}^{6} \Omega_4^4\over \beta_{-1}^3})\;\gamma_5\;{\rm s}\;.
\label{TB}
\end{equation}

Writing the nonthermal electron injection function as
\begin{equation}
{dN_e(\gamma_i,t_i)\over dt_i d\gamma_i} 
\;=\; {\eta\zeta_e L_0 \over m_ec^2 \gamma_w}\;
\delta(\gamma_i-\gamma_w)\;,
\label{dNe}
\end{equation}
the electron Lorentz factor distribution $dN_e(\gamma;t)/d\gamma$ can
be solved to give
\begin{equation}
\gamma^2 {dN_e(\gamma;t)\over d\gamma}\;= \; 
\dot{\cal N}_e T_B \gamma_w^2\;({1\over \hat t}+ 
{1\over \hat \gamma} -1 )^{-2}\;.
\label{g2dNdg}
\end{equation}
In this expression, $\dot{\cal N}_e \equiv \eta \xi_e L_0/(m_ec^2
\gamma_w)$, and we introduce the dimensionless quantities $\hat t = t/
T_B\gamma_w$ and $\hat \gamma = \gamma/ \gamma_w$.  Adiabatic losses,
which are significant on time scales $\sim t/3$, are negligible in
comparison with synchrotron losses of wind electrons when $\hat t \ll
1$, and we restrict ourselves to this regime.

The $\nu L_\nu$ synchrotron radiation flux $\nu L_\nu^{syn} \simeq
{1\over 2} u_B c\sigma_{\rm T} \gamma_s^3 N_e(\gamma_s)$, where
$\gamma_s = \sqrt{\e/\e_B }$, $\e = h\nu/m_ec^2$, $\e_B = B/B_{cr}$,
and the critical magnetic field $B_{cr} = 4.41\times 10^{13}$ G. Thus
\begin{equation}
\nu L_\nu^{syn} (\e ) = c\sigma_{\rm T}\;{B_{cr}^2\over 16\pi}
\;(\dot {\cal N}_e T_B \gamma_w^2)\;({u\over \hat t})^{3/2}\; 
\sqrt{\e }\;({1\over \hat t}+
\sqrt{{u\gamma_w^2\over \e \hat t}}- 1)^{-2}\;\;,
\;\; {\rm for~} \e \leq \e_{max}\;,
\label{nuLnusyn}
\end{equation}
where $u \equiv {\cal B}/\gamma_w B_{cr} T_B \cong 4.5\times 10^{-19}
\beta_{-1}^{3/2}/(\sqrt{\eta\ebw }B_{12} r_{15}^3 \Omega_4^2
\gamma_5$), and
\begin{equation}
\e_{max} \cong \e_B \gamma_w^2 = {{\cal B}
\over B_{cr}}\;{\gamma_w^2\over t} 
\simeq {4.4\times 10^{-9}\over \hat t}\;{\beta_{-1}^{3/2}
\gamma_5\over \sqrt{\eta \ebw } B_{12} r_{15}^3 \Omega_4^2}\;.
\label{emax}
\end{equation}

The synchrotron cooling timescale for wind electrons is $t_{syn} =
\gamma_w T_B \hat t^2$.  In the regime $\hat t \lesssim \beta_0$ where
the wind electrons strongly cool, we can approximate the synchrotron
spectrum by
\begin{equation}
\nu L_\nu^{syn} (\e ) \cong {3\over 8} \xi _e L_0
\;({\e\over \e_{max}})^{1/2}\;H(\e ; \e_0,\e_{max} )\;
\;,
\label{nuLnusynapprox}
\end{equation}
where $H(x;a,b) = 1$ for $a\leq x < b$, and $H=0$ otherwise. The value
of $\hat t$ at $t_{sd}$ is
\begin{equation}
{t_{sd}\over T_B \gamma_w } \;\cong \; 1.4\times 10^{-5} 
\; {j_{0.5}M_3^2\beta_{-1}^3 \over 
\ebw \eta B_{12}^6 r_{15}^{18}\Omega_4^{11} \gamma_5}\;.
\label{ratio}
\end{equation}
Strong cooling holds when $\hat t < t_{sd}/T_B\gamma_w$, noting that
time in physical units can be found from equation (\ref{TB}).

\subsection{Wind Synchrotron Self-Compton Radiation}

The relative importance of synchrotron self-Compton (SSC) to
 synchrotron cooling is given by the ratio $\rho = u_{syn}/u_B$ of the
 synchrotron radiation energy density to $u_B = {\cal B}^2/8\pi t^2$
 (see eq.[\ref{Bwt}]). The PW synchrotron radiation energy density
 $u_{syn}\simeq \kappa (\nu L_\nu^{max})/4\pi R^2 c$, where $\nu
 L_\nu^{max}\cong \xi_e L_0 $ and the parameter $\kappa \geq 1$ is a
 reflection factor such that $\kappa = 1$ corresponds to perfect
 absorption or direct escape (small covering factor) of the PW
 synchrotron radiation, and $\kappa \gg 1$ corresponds to a highly
 reflecting SNR shell with large covering factor.  One obtains $\rho =
 0.033 \kappa \xi_e \beta_{-1}/\ebw \eta$. For large porosity with
 $\kappa \cong 1$ and $\eta \approx 0.1$, we see that the SSC
 contribution is small compared to the synchrotron flux when $\xi_e
 \beta_{-1}/\ebw \lesssim 1$. We restrict ourselves to this regime
 where equation (\ref{gammaw}) is valid.

In the $\delta$-function approximation for the energy gained by a
photon upon being scattered by relativistic electrons\cite{dss97},
with scattering restricted to the Thomson regime, the $\nu L_\nu$ SSC
spectrum from the cooling wind electrons is given by
\begin{equation}
\nu L_\nu^{SSC} \simeq {\sigma_{\rm T}\kappa \xi_e L_0
\over 8\pi v_0^2 T_B }\; {\dot{\cal N}_e \over \hat t^2}\;
\sqrt{{\e\over \e_{max}}}\;\bigl [{1\over a_0(a_0+x)} + a_0^{-2}
 \ln({x\over a_0+x})]|_{x_0}^{x_1},
\label{nuLnuSSC}
\end{equation}
where $a_0 = \hat t^{-1} -1$, $x_0 = \gamma_w \sqrt{\e_0/\e }$, and
$x_1 = \gamma_w \sqrt{\max (\e, \e^{-1}, \e_{max})/\e }$, where we
have used approximation (\ref{nuLnusynapprox}) for the synchrotron
spectrum.

\subsection{Thermal Emission from the Interior of the Shell}

The SNR shell is pictured as having been shredded by the PW and
therefore highly porous, though permeated with small scale ($\sim
10^{12}$-$10^{14}$ cm) density irregularities consisting largely of
metals\cite{dm99,bfd02}.  Note that $0.1 m_{56,-1} M_\odot$ of
$^{56}$Ni provides a radioactive decay power of $\sim 2\times 10^{42}
m_{56,-1}$ ergs s$^{-1}$ before decaying on an $\approx 120$ day
timescale.  The partial covering factor $p_c ~(\leq 1)$ corresponds to
the covering fraction by optically-thick, dense shell
inhomogeneities. Recognizing the severe limitations in the following
expression, we determine the effective temperature $T_{eff}$ in the
interior of the shell through the relation
\begin{equation}
4\pi R^2 \sigma_{\rm SB} T_{eff}^4p_c \cong 
\;{\cal A}p_c\;{\xi_e L_0\over \kappa} + 
2\times 10^{42} m_{56,-1}\exp(-t/120{\rm ~d})\;{\rm ergs~s}^{-1}\;,
\label{Teff}
\end{equation}
where ${\cal A}(\leq 1)$ is an absorption coefficient, dependent in
general on the evolving radiation spectrum, though ${\cal A}$ is here
assumed constant.  The wind heating is more important than heating
from the decay of radioactive Ni when
$({\cal A}/ 0.1) (p_c/ 0.1)\;(\xi_e/ 1/3) \gtrsim$ $ m_{56,-1}
/ B_{12}^2 r_{15}^6 \Omega_4^4$. In this regime,  
\begin{equation}
T_{eff} = ({{\cal A}\xi_e L_0\over 4\pi R^2 \sigma_{\rm
SB}})^{1/4}\approx {825 {\cal A}^{1/4}\over
\sqrt{t/t_{sd}}}\;({\xi_e\over 1/3})^{1/4}\; {B_{12}^{3/2}
r_{15}^{9/2} \Omega_4^{5/3}\over \sqrt{j_{0.5}\beta_{-1}}M_3}\;\;{\rm
K}\;,
\label{Teff1}
\end{equation}
letting $\kappa \approx 1$ because $p_c \sim 0.1$.  For standard
parameters, the shell is heated to $\sim 10^4$ K temperatures when
$t/t_{sd} \ll 1$.  When $B_{12} \gtrsim 4$, this occurs at
$t/t_{sd}\lesssim 1$. The PW heat source is extinguished once the GRB
occurs, and the thermal emission from the SNR shell will decay on the
dynamical timescale $t_{dyn} = R(t)/ c \cong 324
\beta_{-1}\;(j_{0.5}M_e^2/ B_{12}^2 r_{15}^6 \Omega_4^3)\;(t/ t_{sd})$
d, noting that the cooling timescale of a hot shell is generally
shorter than $t_{dyn}$.

The parameters used in Fig.\ 1 are $j_{0.5}=B_{12} = r_{15} =
\Omega_4 = \beta_{-1} = \kappa=M_3 = m_{52,-1} = \gamma_5=1$, $\xi_e =
e_{Bw} = 1/3$, and $p_c = \eta={\cal A} = 0.1$.

\end{document}